\documentclass[twocolumn,secnumarabic,amssymb, nobibnotes, aps, prc, superscriptaddress, nobalancelastpage,longbibliography]{revtex4-2}

\usepackage{CJK}
\usepackage[tbtags]{amsmath}
\usepackage{bm}
\usepackage{graphicx}

\usepackage{tikz,xcolor,hyperref}

\definecolor{lime}{HTML}{A6CE39}
\DeclareRobustCommand{\orcidicon}{
	\begin{tikzpicture}
	\draw[lime, fill=lime] (0,0) 
	circle [radius=0.16] 
	node[white] {{\fontfamily{qag}\selectfont \tiny ID}};
	\draw[white, fill=white] (-0.0625,0.095) 
	circle [radius=0.007];
	\end{tikzpicture}
	\hspace{-2mm}
}
\foreach \x in {A, ..., Z}{%
	\expandafter\xdef\csname orcid\x\endcsname{\noexpand\href{https://orcid.org/\csname orcidauthor\x\endcsname}{\noexpand\orcidicon}}
}

\begin{document}
\begin{CJK*}{UTF8}{gbsn}

\title{Analysis of clustering fragments of $^7$Li and $^7$Be in the microscopic cluster model}

\author{De-Ye~Tao (陶德晔)\orcidA{}}
\affiliation{Key Laboratory of Nuclear Physics and Ion-beam Application (MOE), Institute of Modern Physics, Fudan University, Shanghai 200433, China}

\author{Bo~Zhou (周波)\orcidC{}}
\email{zhou_bo@fudan.edu.cn}
\affiliation{Key Laboratory of Nuclear Physics and Ion-beam Application (MOE), Institute of Modern Physics, Fudan University, Shanghai 200433, China}
\affiliation{Shanghai Research Center for Theoretical Nuclear Physics, NSFC and Fudan University, Shanghai 200438, China}

\author{Si-Min~Wang (王思敏)\orcidB{}}
\affiliation{Key Laboratory of Nuclear Physics and Ion-beam Application (MOE), Institute of Modern Physics, Fudan University, Shanghai 200433, China}
\affiliation{Shanghai Research Center for Theoretical Nuclear Physics, NSFC and Fudan University, Shanghai 200438, China}

\author{Yu-Gang~Ma (马余刚)\orcidD{}}
\email{mayugang@fudan.edu.cn}
\affiliation{Key Laboratory of Nuclear Physics and Ion-beam Application (MOE), Institute of Modern Physics, Fudan University, Shanghai 200433, China}
\affiliation{Shanghai Research Center for Theoretical Nuclear Physics, NSFC and Fudan University, Shanghai 200438, China}

\begin{abstract}
The nuclear structures of $^7$Li~($\alpha+n+n+p$) and $^7$Be~($\alpha+p+p+n$) are studied within the microscopic cluster model, in which the clustering fragments e.g., triton, $^3$He, and even the single nucleons around the core are studied in the $^7$Li and $^7$Be systems. 
We obtain the energy spectra and wave functions of $^7$Li and $^7$Be, and the calculated energy spectra of the ground states and some of the excited states are consistent with experimental data.
To investigate the cluster-formation probabilities, we calculate the reduced-width amplitudes of various binary partitions.
The results show that the $\alpha+t~(^3\mathrm{He})$ cluster structure is dominant in the low-lying states of $^7$Li~($^7$Be), including the ground state and the three lowest excited states.
In some higher states around the single-particle thresholds, the components of $\mathrm{core}+N$ structures, namely $^6\mathrm{Li}+n$ and $^6\mathrm{He}+p$ in $^7$Li and $^6\mathrm{Li}+p$ in $^7$Be, are significantly enhanced.
\end{abstract}

\maketitle

\section{Introduction}
\label{intro}

Nuclear clustering is one of the most intriguing phenomena and has drawn extensive studies aiming to understand and explore the cluster structures in the ground and various excited states of nuclei~\cite{freer2018,liu2018,zhou2019FP,Ma_NT,Ma_SCPMA}. 
Beyond nuclear structure, clustering plays important roles in many research fields.
In recent years, by taking into account the clustering effects, a wide range of studies have been promoted in the fields of nucleosynthesis~\cite{sun2024,pais2018}, photonuclear reactions using laser-electron gamma sources \cite{Huang2017PRC,Huang2020PRC,NST1,NST2,NST3}, heavy-ion collisions~\cite{cao2023,shi2021,ma2020,xu2018,ZhangYX,CaoYT,WangSS}, and the equation of state (EOS) of nuclear matter~\cite{tanaka2021,ropke2011}.

The cluster structures and related reactions of $^7$Li and $^7$Be nuclei are essential for understanding the primordial abundances of lithium isotopes, which remains an unsolved problem in astrophysics~\cite{bertulani2016,cyburt2016,fields2011}.
Using various microscopic cluster models~\cite{freer1997,saito1977}, low-lying states of $^7$Li~\cite{mihailovic1978,beck1981,fujiwara1983,sharma1984,toshitaka1984,fujiwara1985,fujiwara1991,kanada1995,arai2001,furumoto2018} and $^7$Be~\cite{descouvemont2004,arai2002,descouvemont1995,descouvemont1994a,descouvemont1994b,descouvemont1988} have been intensively investigated for decades. 
A diversity of cluster structures exist in $^7$Li and $^7$Be, among which the structures of $\alpha+t$ and $\alpha+{}^3\mathrm{He}$ are the most well-known cluster configurations, respectively.
The $\alpha$ cluster, as the most stable subunit, is widely used as the building block for various microscopic cluster models~\cite{brink1966,horiuchi1977,zhou2023,zhou2019,zhou2014}.
On the other hand, the formation of triton and $^3$He clusters in nuclei has recently drawn increasing attention both experimentally~\cite{ma2021,kawabata2007,nakayama2004} and theoretically~\cite{zhou2018,suhara2012,enyo2007}.
Thus, studying the emergence of triton and $^3\mathrm{He}$ clusters in $^{7}$Li and $^{7}$Be nuclei, as simple examples, would be informative for further studies of these cluster formations in heavier and more complex nuclei.

Distinctly from the $\alpha$ cluster, triton and $^3$He possess rather low binding energies, which means that they are easy to break up and, by interacting with the $\alpha$ cluster, form more complicated multi-cluster configurations, including $^6$Li, $^6$He, $d$, and single nucleons~\cite{he2020,liu2020,HuangBS_Li6}. 
For example, in the higher excited states of $^{7}$Li and $^{7}$Be, the $\mathrm{core}+N$ structures have attracted much current interest~\cite{dubovichenko2022,kiss2021,piatti2020,vorabbi2019,dong2017,li2018,he2013}. 
In recent years, with the improvement of experimental techniques, abundant evidence of cluster configurations, including the traditional $\alpha+t~(^3\mathrm{He})$ structure and various novel $\mathrm{core}+N$ structures, has been observed in $^{7}$Li and $^{7}$Be~\cite{chattopadhyay2018,pakou2016,shrivastava2013}.
Theoretically, a recent no-core shell model~(NCSM) calculation~\cite{vorabbi2019} predicted that a near-threshold $^{6}\mathrm{He}+p$ resonance exists in $^{7}$Li, while the analogous $^{6}\mathrm{Li}+p$ structure in $^{7}$Be is not found. 

The aims of the present work are to quantify the components of various cluster configurations, as well as try to verify the results obtained by the NCSM in Ref.~\cite{vorabbi2019} using the microscopic cluster framework.
We apply the four-body generator coordinate method~(GCM) to study the clustering structure of the $^7$Li and $^7$Be systems.
The cluster configurations of the wave functions obtained from the GCM are evaluated by calculating the reduced-width amplitude~(RWA)~\cite{tao2024}, defined in the $R$-matrix theory~\cite{Descouvemont2010} as the probability of cluster formation at a given distance. 
The calculations of RWA have been employed in a wide range of microscopic studies to estimate the various cluster weights in nuclei~\cite{descouvemont2022,zhao2022,lyu2018,zhou2016,yamada2015,enyo2014,arai2003}.

The article is arranged as follows. 
In Sec.~\ref{model} we briefly present the theoretical framework of the GCM and the calculation of RWA. 
The calculated results and discussions regarding the cluster structures of $^7$Li and $^7$Be are presented in Sec.~\ref{results}. 
Finally, we summarize the main conclusions and discuss potential further studies in Sec.~\ref{summary}.

\section{Model Description}
\label{model}

According to the framework of the GCM, the total wave function of $^{7}$Li~($^{7}$Be) can be written as the superposition of angular-momentum-projected and parity-projected Brink wave functions
\begin{equation}
    \Psi^{J\pi}_M = \sum_{i,K}c_{i,K} \hat P_{MK}^J \hat P^\pi \Phi^\mathrm{B}(\{\boldsymbol R\}_i),
\label{wf_tot}
\end{equation}
in which $\hat P_{MK}^J$ and $\hat P^\pi$ are the angular-momentum and parity projectors, respectively.
The index $i$ indicates a specified set of generator coordinates $\{\bm R_1,\cdots, \bm R_4\}$. 
The Brink wave function is fully antisymmetrized as follows: 
\begin{equation}
\begin{aligned}
    \Phi^\mathrm{B}(\bm R_1,\cdots,\bm R_4)=\mathcal{A}[&\phi_1(\bm{R}_1)\cdots\phi_4(\bm{R}_1) \cr
    \times{}&\phi_5(\bm{R}_2)\phi_6(\bm{R}_3)\phi_7(\bm{R}_4)],
\end{aligned}
\end{equation}
where the wave function of the $k$-th nucleon is defined as a Gaussian wave packet
\begin{equation}
    \phi_k(\bm R_j)=\frac{1}{(\pi b^2)^{3/4}}\exp\left[-\frac{1}{2b^2}(\bm{r}_k-\bm{R}_j)^2\right] \chi_k \tau_k .
\end{equation}

In the present calculation, the oscillator parameter for the single-particle wave functions is set to $b=1.46~\mathrm{fm}$.
The Hamiltonian of the system includes kinetic, central N-N, spin-orbit, and Coulomb parts
\begin{equation}
\begin{aligned}
    \hat H={}&-\frac{\hbar^2}{2m}\sum_i \nabla_i^2 - T_\mathrm{c.m.} \cr 
    &+ \sum_{i<j}\hat V_{ij}^\mathrm{NN} + \sum_{i<j}\hat V_{ij}^\mathrm{LS} + \sum_{i<j}\hat V_{ij}^\mathrm{C}.
\label{ham}
\end{aligned}
\end{equation}
The Volkov No.2 potential~\cite{volkov1965} is taken as the central N-N potential
\begin{equation}
    \hat V_{ij}^\mathrm{NN}=\sum_{n=1}^2v_ne^{-\frac{r_{ij}^2}{a_n^2}}(W+B\hat P_\sigma-H\hat P_\tau-M\hat P_\sigma \hat P_\tau)_{ij}
\end{equation}
with $a_1={1.01}~\mathrm{fm}$, $a_2={1.8}~\mathrm{fm}$, $v_1={61.14}~\mathrm{MeV}$, $v_2={-60.65}~\mathrm{MeV}$, $W=1-M$, $M=0.6$ and $B=H=0.125$.
The G3RS potential~\cite{tamagaki1968,yamaguchi1979} is used for the spin-orbit term
\begin{equation}
    \hat V_{ij}^\mathrm{LS}=v_0(e^{-d_1r_{ij}^2}-e^{-d_2r_{ij}^2})\hat P(^3O)\bm L\cdot\bm S,
\end{equation}
where $\hat P(^3O)$ is the projection operator onto a triplet odd state, the strength $v_0=2000~\mathrm{MeV}$, and the parameters $d_1$ and $d_2$ are set to $5.0~\mathrm{fm}^{-2}$ and $2.778~\mathrm{fm}^{-2}$, respectively.
The coefficients \{$c_{i,K}$\} in Eq.~(\ref{wf_tot}) are determined by solving the Hill-Wheeler equation.
To estimate the energies of the resonance states, the radius constraint method (RCM)~\cite{funaki2006} is applied in the GCM calculation.
The radius cutoff parameter is set to $R_{\mathrm{cut}}=5~\mathrm{fm}$.

The obtained GCM wave functions for the ground and excited states are used to analyze the cluster structures of the system.
By employing the Laplace expansion method~\cite{chiba2017}, we calculate the reduced-width amplitudes, defined as~\cite{Descouvemont2010}
\begin{widetext}
\begin{equation}
    y_{j_1\pi_1j_2\pi_2j_{12}}^{J\pi}(a)=\sqrt{\frac{A!}{(1+\delta_{C_1C_2})C_1!C_2!}}
        \left<\frac{\delta(r-a)}{r^2}\left[Y_l(\hat{\bm r})\otimes
        \left[\Phi_{C_1}^{j_1\pi_1}\otimes\Phi_{C_2}^{j_2\pi_2}\right]_{j_{12}}\right]_{JM}\middle|\Psi^{J\pi}_M\right> ,
\label{rwa}
\end{equation}
\end{widetext}
which quantify the probabilities of forming various two-cluster structures at specified distances.

\section{Results and Discussion}
\label{results}

\begin{figure*}[htpb]
    \centering
    \includegraphics[width=\linewidth]{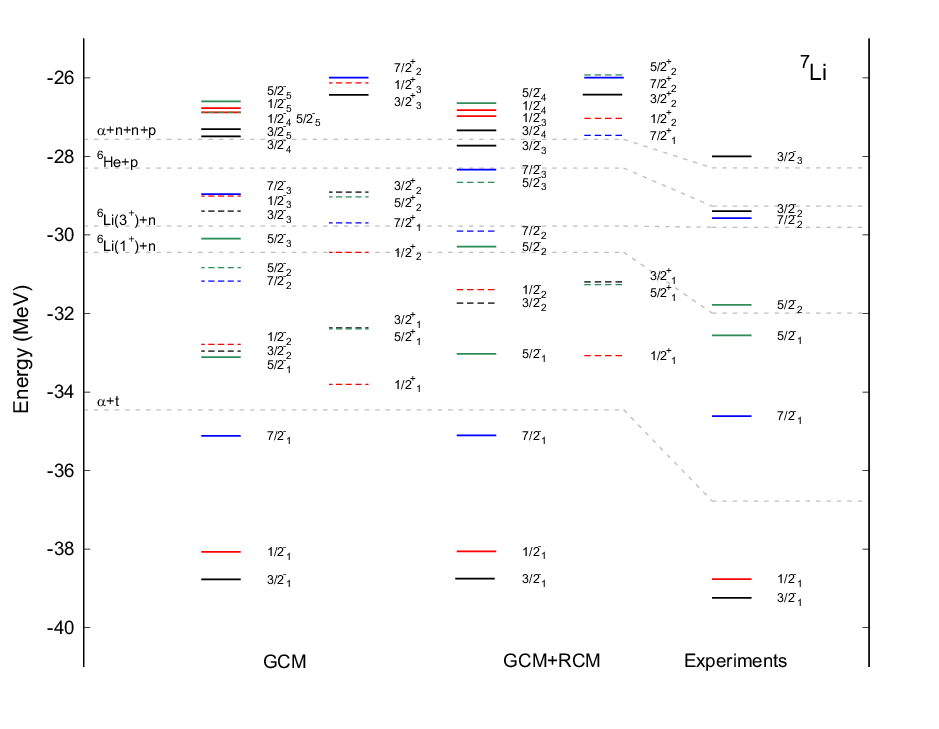}
    \caption{The calculated energy spectra of $^7$Li compared with the experimental values~\cite{tilley2002}. The states with solid lines are physical states according to our analysis, and the states with dashed lines may contain large components of non-physical continuum states. The gray dashed lines represent thresholds of various cluster structures.}
    \label{lev_Li7}
\end{figure*}

\begin{figure*}[htpb]
    \centering
    \includegraphics[width=\linewidth]{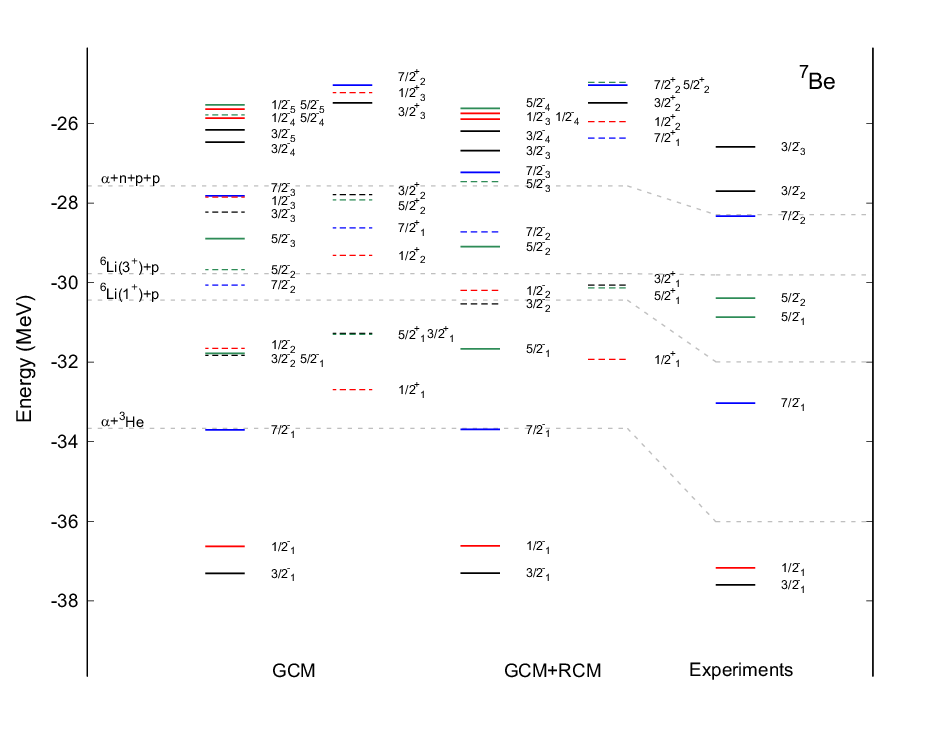}
    \caption{The calculated energy spectra of $^7$Be compared with the experimental values~\cite{tilley2002}. The states with solid lines are physical states according to our analysis, and the states with dashed lines may contain large components of non-physical continuum states. The gray dashed lines represent thresholds of various cluster structures.}
    \label{lev_Be7}
\end{figure*}

The calculated energy levels are shown in Fig.~\ref{lev_Li7} for $^7$Li and Fig.~\ref{lev_Be7} for $^7$Be. 
In these Figures, the first two columns from the left display the results from the GCM calculation without applying the RCM, for both negative and positive parity states, while the third and fourth columns present the results obtained using the RCM.
The experimental data are shown in the right of the Figures.
From these Figures, we can see that the lowest four states of $^7$Li and $^7$Be are well reproduced by our calculations. 
In order to test the obtained wave functions of the ground and lowest excited states, we calculate the root-mean-square (r.m.s.) radii for the ground states, as well as the electromagnetic transition strengths between various states of $^7$Li and $^7$Be.
The calculated r.m.s.\ charge radius of $^7$Li is $2.43~\mathrm{fm}$, which is quite consistent with the experimental value of $2.42~\mathrm{fm}$~\cite{sanchez2006}. 
Meanwhile, the charge radius of $^7$Be is obtained to be $2.63~\mathrm{fm}$. 
The calculated and experimental electromagnetic transition strengths are listed in Table~\ref{obs}.
The theoretical results of $B(E2;3/2^-\to 1/2^-)$ for $^7$Li and $^7$Be, namely $8.91~e^2\mathrm{fm}^4$ and $26.2~e^2\mathrm{fm}^4$, coincide well with the experimental values of $8.3~e^2\mathrm{fm}^4$~\cite{tilley2002} and $26~e^2\mathrm{fm}^4$~\cite{henderson2019}, respectively.
For transitions among higher excited states, we list some of the predicted values based on our calculations.

Compared with the lowest four states, the calculated higher states are more complicated and not good consistent with experimental data.
One reason for this discrepancy is that continuum states are inevitably included during the GCM calculations. 
For simplicity, to remove the non-physical continuum states, we apply the RCM~\cite{funaki2006} and the obtained spectra are shown to the right of the original spectra.
The first four states remain almost unchanged after applying the RCM, as expected.
However, in the higher energy range, some states change dramatically or disappear after using the RCM, as indicated by the dashed lines.
This suggests a scattering-like feature with large components of continuum states involved in these original states, as they exhibit significant sensitivity to the radius.
Consequently, it is difficult to discuss these states within the framework of our current model. 
On the other hand, some states, even with high excitation energies, change only slightly after applying the RCM, as shown in Fig.~\ref{lev_Li7} and Fig.~\ref{lev_Be7} with solid lines.
According to the level order in the spectra obtained by GCM with RCM, these states are denoted as $5/2^-_2$, $7/2^-_3$, $3/2^-_3$, $3/2^-_4$, $1/2^-_3$, $1/2^-_4$, $5/2^-_4$, $3/2^+_2$, and $7/2^+_2$.
Notably, the first four of these states, i.e., $5/2^-_2$, $7/2^-_3$, $3/2^-_3$, and $3/2^-_4$, are basically consistent with experimental data for the energy level order.
In this case, these selected states (solid lines), together with the lowest four states, represent the physical states obtained in our model, and the subsequent discussion will focus on these states.

\begin{table}
\centering
\caption{The calculated $B(E2)$ transition strengths ($e^2\mathrm{fm}^4$) of ground and excited states of $^7$Li and $^7$Be.}
\begin{tabular}{lcccc}
    \hline\hline
                           & $^7$Li & Exp.  & $^7$Be & Exp. \\
    \hline
    $B(E2;3/2^-_1\to 1/2^-_1)$ & $8.91$ & $8.3$ & $26.2$ & $26$ \\
    $B(E2;3/2^-_1\to 7/2^-_1)$ & $15.1$ & $6.8$ & $48.2$ &      \\
    $B(E2;3/2^-_1\to 5/2^-_1)$ & $3.89$ &       & $11.3$ &      \\
    $B(E2;1/2^-_1\to 5/2^-_1)$ & $36.2$ &       & $82.3$ &      \\
    $B(E2;7/2^-_1\to 5/2^-_1)$ & $0.663$&       & $4.21$ &      \\
    \hline
\end{tabular}
\label{obs}
\end{table}

\begin{figure*}[htpb]
    \centering
    \includegraphics[width=\linewidth]{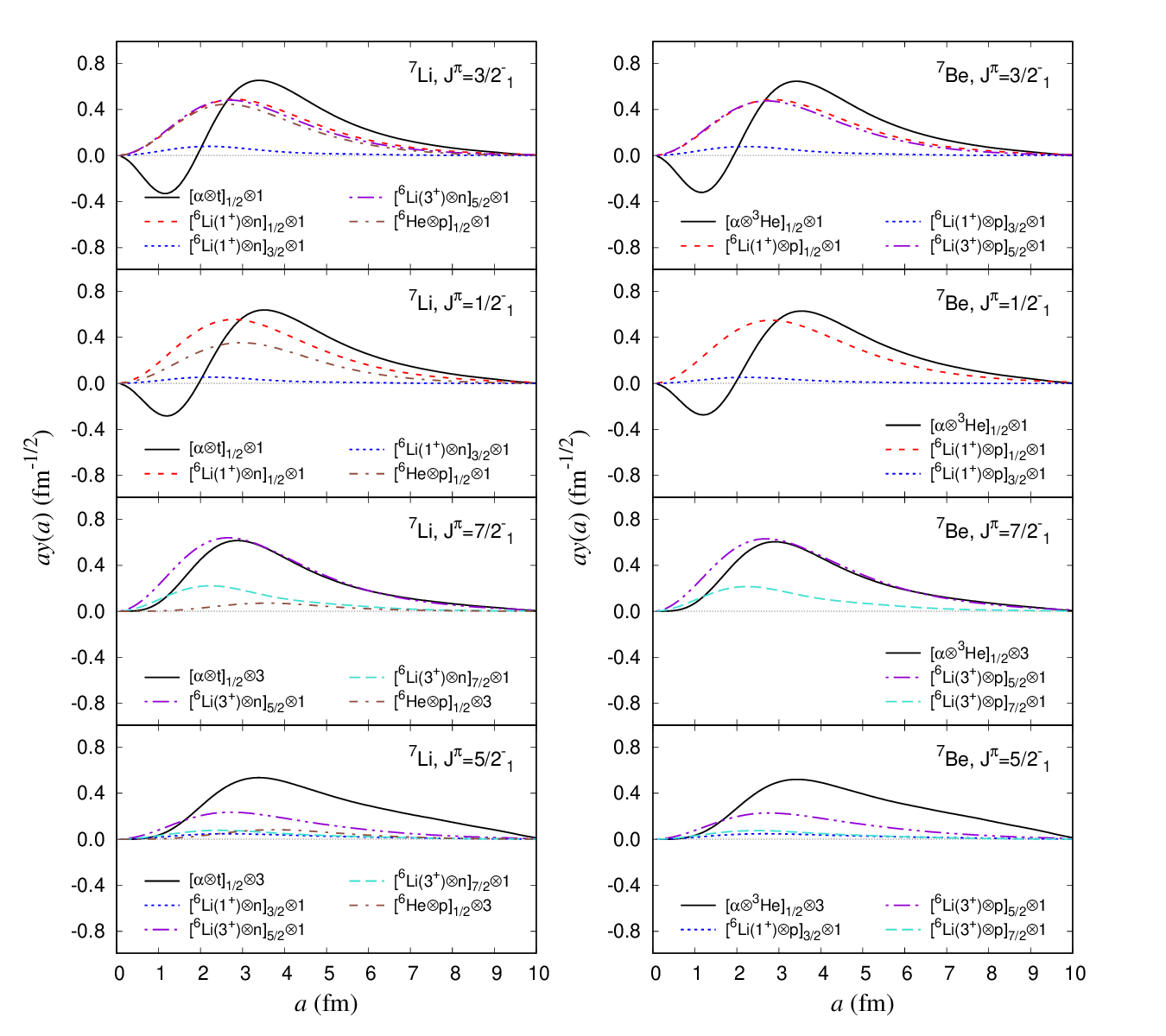}
    \caption{The calculated RWA for two-body clustering structure in different channels for the $3/2^-_1$, $1/2^-_1$, $7/2^-_1$, and $5/2^-_1$ states of $^7$Li and $^7$Be.}
\label{rwa_1}
\end{figure*}

Since we obtained the relatively high-energy states for $^7$Li and $^7$Be, it is interesting to explore their clustering structures. 
Using Eq.~(\ref{rwa}), we calculate the RWA of two-cluster structures including $\alpha+t$, $^6\mathrm{Li}(1^+,3^+)+n$, and $^6\mathrm{He}+p$ for $^7$Li, and $\alpha+{^3\mathrm{He}}$ and $^6\mathrm{Li}(1^+,3^+)+p$ for $^7$Be.
The RWA results for the $3/2^-_1$, $1/2^-_1$, $7/2^-_1$, and $5/2^-_1$ states of $^7$Li and $^7$Be are shown in Fig.~\ref{rwa_1}.
To identify the coupling channel, we use the expression $[C_1(j_1)\otimes C_2(j_2)]_{j_{12}}\otimes l$.
Here, the $C_1$ cluster with angular momentum $j_1$ and the $C_2$ cluster with angular momentum $j_2$ are coupled to the angular momentum of $j_{12}$, which is then coupled with the orbital angular momentum $l$ of the relative motion between the clusters to form the total angular momentum of the nucleus.
It should be noted that, due to the RCM, the amplitudes reach zero when the distance between clusters exceeds $10~\mathrm{fm}$. 
The results clearly show that in the four lowest states, i.e., $3/2^-_1$, $1/2^-_1$, $7/2^-_1$, and $5/2^-_1$, of $^7$Li, the configurations of $\alpha+t$ are significant.
For the $3/2^-_1$ and $1/2^-_1$ states, the RWA of $\alpha+t$ have one node around $2~\mathrm{fm}$ due to the Pauli forbidden states, while for the $7/2^-_1$ and $5/2^-_1$ states have none node.
The $5/2^-_1$ state, as a resonance state with relatively high excitation energy, exhibits a long tail in the $\alpha+t$ RWA, which might be an intrinsic character of this state or caused by the $\alpha+t$ continuum, even with a relatively large orbital angular momentum $l$.
Apart from the $\alpha+t$ components, the $^6\mathrm{He}+p$ and $^6\mathrm{Li}+n$ configurations are also non-negligible in the ground state $3/2^-_1$ and the first excited state $1/2^-_1$ of $^7$Li, and these components are competitive with each other. 
The component of $^6\mathrm{He}+p$ in the $3/2^-_1$ state is slightly larger than that in the $1/2^-_1$ state.
The $^6\mathrm{Li}(1^+,3^+)+n$ configurations are more complicated as the result of the variety of angular-momentum-coupling channels.
In the ground state, the $[^6\mathrm{Li}(1^+)\otimes n]_{1/2}\otimes1$ channel is much more significant than the $[^6\mathrm{Li}(1^+)\otimes n]_{3/2}\otimes1$ channel. 
Meanwhile, the channel composed of the first excited state of $^6$Li, $[^6\mathrm{Li}(3^+)\otimes n]_{5/2}\otimes1$, also exists in the ground state of $^7$Li, with an amplitude comparable with that of $[^6\mathrm{Li}(1^+)\otimes n]_{1/2}\otimes1$.
Notably, in the ground state of $^7$Li, the amplitudes of the three main $\mathrm{core}+N$ channels, namely $[^6\mathrm{He}\otimes p]_{1/2}\otimes1$, $[^6\mathrm{Li}(1^+)\otimes n]_{1/2}\otimes1$, and $[^6\mathrm{Li}(3^+)\otimes n]_{5/2}\otimes1$, are quite similar when $a$ is small.
The $^6\mathrm{Li}(1^+)+n$ components in the first excited state possess similar features as in the ground state.
The channel of $[^6\mathrm{Li}(1^+)\otimes n]_{1/2}\otimes1$ constitutes the main component and is slightly larger than that in the ground state.
In the $7/2^-_1$ state, $[^6\mathrm{Li}(3^+)\otimes n]_{5/2}\otimes1$ is the most dominant $\mathrm{core}+N$ configuration, comparable with the $\alpha+t$ configuration in amplitude. 
It should be noted that the $7/2^-_1$ state is located about $5~\mathrm{MeV}$ below the $^6\mathrm{Li}(3^+)+n$ threshold.
The other $\mathrm{core}+N$ channels, including $[^6\mathrm{Li}(3^+)\otimes n]_{7/2}\otimes1$ and $[^6\mathrm{He}\otimes p]_{1/2}\otimes3$, exhibit minor amplitudes in this state. 
For the $5/2^-_1$ state, the channel of $[^6\mathrm{Li}(3^+)\otimes n]_{5/2}\otimes1$ has the largest amplitude among various $\mathrm{core}+N$ configurations. 
However, this component is still insignificant compared to the predominant $\alpha+t$ structure.
A notable feature of these four lowest states is that in the clustering channels of $^6\mathrm{Li}(1^+,3^+)+n$, $^6$Li and neutron tend to couple to a lower $j_{12}$ angular momentum.
Specifically, for the $^6\mathrm{Li}(1^+)\otimes n$ coupling, $j_{12}$ is more likely to be $1/2$, while for the $^6\mathrm{Li}(3^+)\otimes n$ coupling, $j_{12}$ is more likely to be $5/2$.

\begin{figure*}[htpb]
    \centering
    \includegraphics[width=\linewidth]{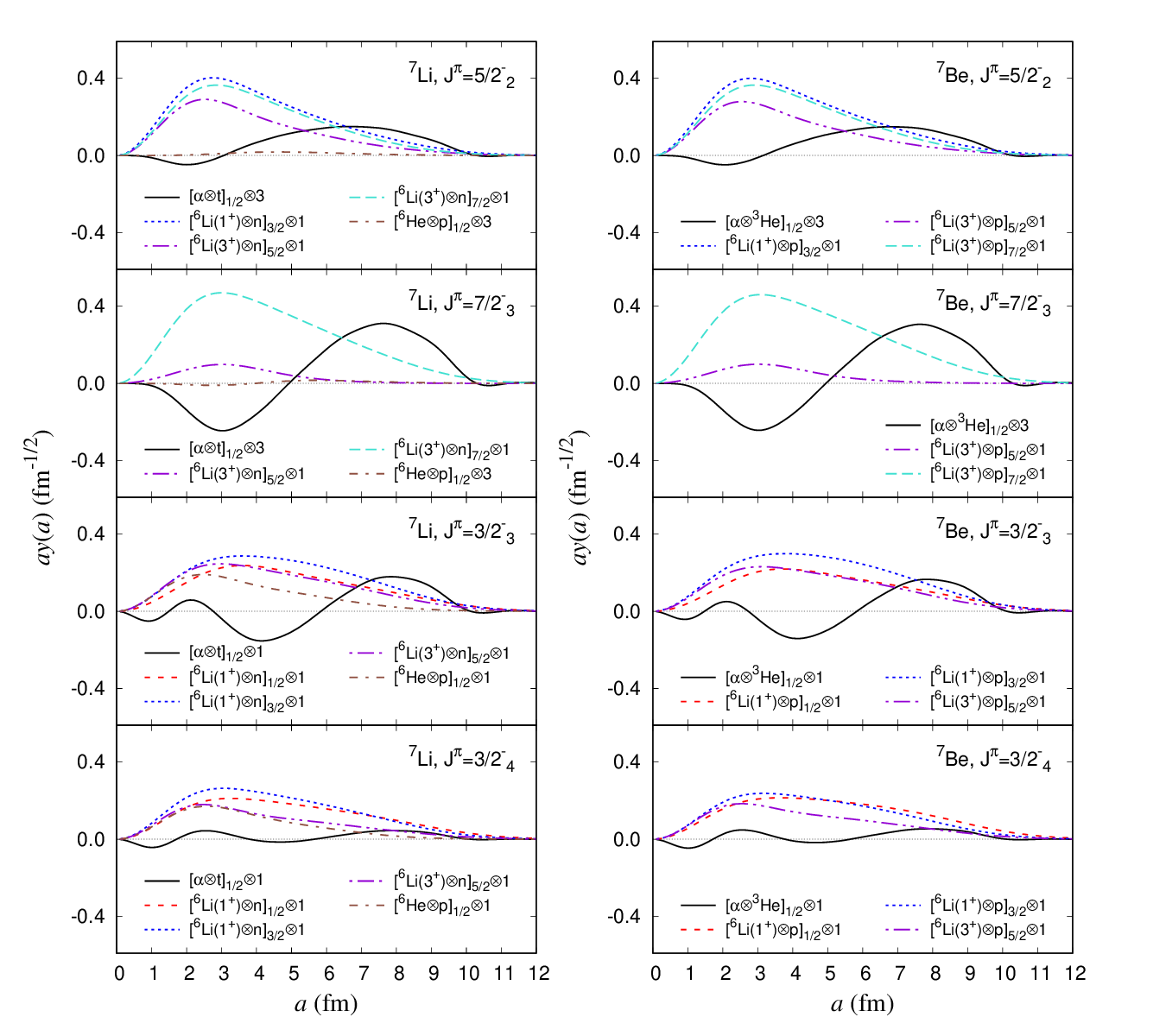}
    \caption{The calculated RWA for two-body clustering structure in different channels for the $5/2^-_2$, $7/2^-_3$, $3/2^-_3$, and $3/2^-_4$ states of $^7$Li and $^7$Be.}
\label{rwa_2}
\end{figure*}

For the higher-lying four states that may correspond to the experimental levels, namely $5/2^-_2$, $7/2^-_3$, $3/2^-_3$, and $3/2^-_4$, as shown in Fig.~\ref{rwa_2}, the components of $\alpha+t$ are reduced, and the $\mathrm{core}+N$ structures dominate, especially for small $a$ values.
This can be understood considering that these states are far from the $\alpha+t$ threshold and closer to the $^6\mathrm{Li}+n$ and $^6\mathrm{He}+p$ thresholds.
This phenomenon has been discussed as the ``alignment'' effect~\cite{okolowicz2020}, where narrow resonances tend to be found in the vicinity of particle decay thresholds.
Besides the smaller amplitudes of the $\alpha+t$ configuration compared to the lowest four states, the $\alpha+t$ RWA for these higher states exhibit more nodes, and the positions of maxima are shifted to larger $a$ values.
Notably, in the $7/2^-_3$ state, the $\alpha+t$ structure is comparable with the dominant $[^6\mathrm{Li}(3^+)\otimes n]_{7/2}\otimes1$ configuration.
However, since the maximum of this $\alpha+t$ RWA is reached at a very large distance ($\approx8~\mathrm{fm}$), such a component could primarily be attributed to the non-physical continuum.
In contrast, the $\mathrm{core}+N$ configurations are much more prominent for small $a$ values. 
The $5/2^-_2$ state is dominated by various $^6\mathrm{Li}(1^+,3^+)+n$ configurations, including the competitive $[^6\mathrm{Li}(1^+)\otimes n]_{3/2}\otimes1$ and $[^6\mathrm{Li}(3^+)\otimes n]_{7/2}\otimes1$ channels, as well as the smaller $[^6\mathrm{Li}(3^+)\otimes n]_{5/2}\otimes1$ channel.
In the $7/2^-_3$ state, the $[^6\mathrm{Li}(3^+)\otimes n]_{7/2}\otimes1$ configuration is the most significant one.
The $3/2^-_3$ and $3/2^-_4$ states, with the same angular momentum and parity, show similar patterns in their $\mathrm{core}+N$ RWA.
The three $^6\mathrm{Li}+n$ channels, namely $[^6\mathrm{Li}(1^+)\otimes n]_{1/2}\otimes1$, $[^6\mathrm{Li}(1^+)\otimes n]_{3/2}\otimes1$, and $[^6\mathrm{Li}(3^+)\otimes n]_{5/2}\otimes1$, exhibit similar zero-node and long-tail RWA curves.
In addition to $^6\mathrm{Li}+n$, the $^6\mathrm{He}+p$ structure is also present in these two $3/2^-$ states, with amplitude secondary to those of the $^6\mathrm{Li}+n$ channels.
One distinction between the $3/2^-_3$ and $3/2^-_4$ states is that the peaks of $^6\mathrm{Li}+n$ RWA in the $3/2^-_4$ state are shifted inward compared to those in the $3/2^-_3$ state.

\begin{table*}
\centering
\caption{The calculated spectroscopic factors of $3/2^-_1$, $1/2^-_1$, $7/2^-_1$, and $5/2^-_1$ states of $^7$Li and $^7$Be.}
\begin{tabular*}{\linewidth}{@{\extracolsep{\fill}}ccccc}
    \hline\hline
              & $^7$Li     & SF & $^7$Be                 & SF \\
    \hline
    $3/2^-_1$   & $\alpha+t$ & $1.03$    
                & $\alpha+{}^3\mathrm{He}$ & $1.02$    \\
                & $[^6\mathrm{Li}(1^+)\otimes n]_{1/2}\otimes1$ & $0.657$     
                & $[^6\mathrm{Li}(1^+)\otimes p]_{1/2}\otimes1$ & $0.660$     \\
                & $[^6\mathrm{Li}(1^+)\otimes n]_{3/2}\otimes1$ & $1.26\times10^{-2}$ 
                & $[^6\mathrm{Li}(1^+)\otimes p]_{3/2}\otimes1$ & $1.23\times10^{-2}$  \\
                & $[^6\mathrm{Li}(3^+)\otimes n]_{5/2}\otimes1$ & $0.610$     
                & $[^6\mathrm{Li}(3^+)\otimes p]_{5/2}\otimes1$ & $0.601$  \\
                & $^6\mathrm{He}+p$ & $0.509$    &   &    \\     
    $1/2^-_1$   & $\alpha+t$ & $1.02$    
                & $\alpha+{}^3\mathrm{He}$ & $1.01$    \\
                & $[^6\mathrm{Li}(1^+)\otimes n]_{1/2}\otimes1$ & $0.855$     
                & $[^6\mathrm{Li}(1^+)\otimes p]_{1/2}\otimes1$ & $0.853$    \\
                & $[^6\mathrm{Li}(1^+)\otimes n]_{3/2}\otimes1$ & $5.86\times10^{-3}$     
                & $[^6\mathrm{Li}(1^+)\otimes p]_{3/2}\otimes1$ & $5.62\times10^{-3}$   \\
                & $^6\mathrm{He}+p$ & $0.337$  &    &    \\     
    $7/2^-_1$   & $\alpha+t$ & $0.935$   
                & $\alpha+{}^3\mathrm{He}$ & $0.922$   \\
                & $[^6\mathrm{Li}(3^+)\otimes n]_{5/2}\otimes1$ & $1.13$     
                & $[^6\mathrm{Li}(3^+)\otimes p]_{5/2}\otimes1$ & $1.13$    \\
                & $[^6\mathrm{Li}(3^+)\otimes n]_{7/2}\otimes1$ & $0.110$     
                & $[^6\mathrm{Li}(3^+)\otimes p]_{7/2}\otimes1$ & $0.105$    \\
    $5/2^-_1$   & $\alpha+t$ & $0.911$   
                & $\alpha+{}^3\mathrm{He}$ & $0.900$   \\
                & $[^6\mathrm{Li}(3^+)\otimes n]_{5/2}\otimes1$ & $0.161$     
                & $[^6\mathrm{Li}(3^+)\otimes p]_{5/2}\otimes1$ & $0.157$    \\
                & $[^6\mathrm{Li}(3^+)\otimes n]_{7/2}\otimes1$ & $1.54\times10^{-2}$ 
                & $[^6\mathrm{Li}(3^+)\otimes p]_{7/2}\otimes1$ & $1.50\times10^{-2}$   \\

    \hline
\end{tabular*}
\label{s-factor}
\end{table*}

For the states mentioned above, the comparisons between the RWA of $^7$Li and $^7$Be are particularly interesting.
As mirror nuclei, $^7$Li and $^7$Be exhibit similar features in the RWA of analogous configurations.
The RWA curves of $\alpha+t$ in $^7$Li and $\alpha+{}^3$He in $^7$Be for each corresponding state are quite similar, except that the curves for states of $^7$Be have slightly longer tails, likely due to the difference between the Coulomb interactions in $^7$Li and $^7$Be.
These characteristics are also observed in the $\mathrm{core}+N$ RWA for analogous $^6\mathrm{Li}+n$ and $^6\mathrm{Li}+p$ configurations.

As important quantities for measuring cluster components and input parameters in evaluating cross sections for nuclear reactions~\cite{Descouvemont2010,enyo2014,chiba2017}, the spectroscopic factors (SF), defined as
\begin{equation}
    S=\int_0^\infty da \left|a y(a)\right|^2 ,
\end{equation}
are calculated and listed in Table~\ref{s-factor} for the lowest four states of $^7$Li and $^7$Be.
The calculated SF for the $\alpha+t$ components in the ground and first excited states of $^7$Li are $1.03$ and $1.02$, respectively, which are consistent with the values of $1.10$ and $1.05$ obtained by the resonant group method (RGM) calculation~\cite{descouvemont2022}. 
In the RGM calculation, the breakup effect of the triton cluster is considered by coupling the channels of $\alpha+t$, $^6\mathrm{He}+p$, and $^6\mathrm{Li}+n$.
In contrast, our GCM calculations assume the $\alpha+n+n+p$ four-body cluster model.
For the $^6\mathrm{He}+p$ structure, the calculated SF are $0.51$ for the ground state and $0.34$ for the $1/2^-_1$ state, which are slightly overestimated compared to the RGM results of $0.41$ and $0.22$, respectively.
Similarly, our SF results for the $^6\mathrm{Li}(1^+)+n$ structure are $0.67$ for the ground state and $0.86$ for the $1/2^-$ state, also slightly overestimated compared to the RGM values of $0.53$ for the ground state and $0.79$ for the $1/2^-_1$ state. 
Notably, the calculated $^6\mathrm{Li}(1^+)+n$ SF of $0.67$ for the ground state is quite close to the \textit{ab initio} result of $0.68$~\cite{timofeyuk2009}.
Consistently with the RWA results, the calculated SF for $^7$Be are similar to those for $^7$Li for the corresponding channels.

\begin{figure*}[htpb]
    \centering
    \includegraphics[width=\linewidth]{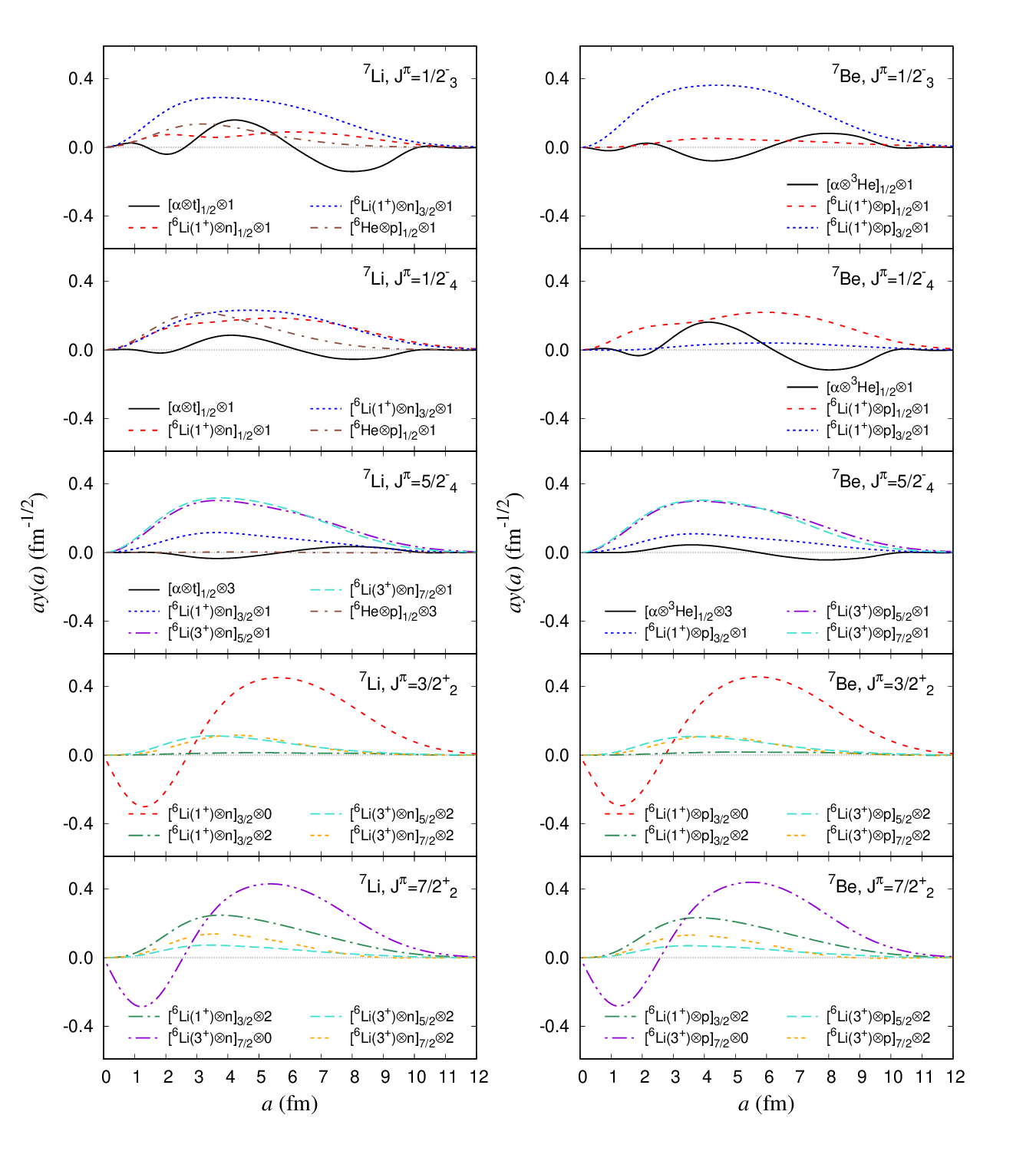}
    \caption{The calculated RWA for two-body clustering structure in different channels for the $1/2^-_3$, $1/2^-_4$, $5/2^-_4$, $3/2^+_2$, and $7/2^+_2$ states of $^7$Li and $^7$Be.}
\label{rwa_3}
\end{figure*}

Next, we analyze the cluster structure for some states not yet observed in experiments, lying high above the $\alpha+n+n(p)+p$ threshold.
The RWA results are shown in Fig.~\ref{rwa_3}.
In all these five states, the $\alpha+t~(^3\mathrm{He})$ channels are drastically suppressed, while the $\mathrm{core}+N$ structures are predominant.
In the $1/2^-_3$ state of $^7$Li, the channel of $[^6\mathrm{Li}(1^+)+n]_{3/2}\otimes1$ accounts for the largest percentage and dominates.
For the $1/2^-_4$ state, the main configurations are $[^6\mathrm{Li}(1^+)\otimes n]_{1/2}\otimes1$, $[^6\mathrm{Li}(1^+)\otimes n]_{3/2}\otimes1$, and $[^6\mathrm{He}\otimes p]_{1/2}\otimes1$, which exhibit analogous zero-node RWA with competitive amplitudes.
Notably, although the $^6\mathrm{He}+p$ structure is comparable with $^6\mathrm{Li}(1^+)+n$ configurations in the small $a$ range, the curves decrease more rapidly for larger $a$ values.
For $^7$Be, the two $^6\mathrm{Li}(1^+)+p$ channels are much more divergent from each other in these two $1/2^-$ states.
The $[^6\mathrm{Li}(1^+)\otimes p]_{3/2}\otimes1$ and $[^6\mathrm{Li}(1^+)\otimes p]_{1/2}\otimes1$ are exclusively predominant in $1/2^-_3$ and $1/2^-_4$, respectively.
This indicates that for the $1/2^-_3$ state, the $^6$Li-proton-coupled angular momentum is more likely to be $3/2$, while for the $1/2^-_4$ state, it prefers to be $1/2$.
The reason may be that, with higher excitation energy above the threshold, the low angular momentum enhances the influence of the continuum due to the lower centrifugal barrier~\cite{wang2021,wang2022,mao2020}, which will introduce more coupling to the environment and thus break the mirror symmetry~\cite{wang2019,pfutzner2023,zhang2022}. 
This effect is also known as the Thomas-Ehrman shift~\cite{thomas1951,ehrman1951}.
In the $5/2^-_4$ states of $^7$Li and $^7$Be, the channels of $[^6\mathrm{Li}(3^+)+n(p)]_{5/2}\otimes1$ and $[^6\mathrm{Li}(3^+)+n(p)]_{7/2}\otimes1$ comprise the main components, with amplitudes close to each other.
Distinctively, in both $^7$Li and $^7$Be, our predicted positive-parity states, $3/2^+_2$ and $7/2^+_2$, exhibit significant one-node $\mathrm{core}+N$ RWA curves for $[^6\mathrm{Li}(1^+)+n(p)]_{3/2}\otimes0$ and $[^6\mathrm{Li}(3^+)+n(p)]_{7/2}\otimes0$, respectively, which are quite similar to each other.
Compared to these two channels, the other $\mathrm{core}+N$ channels are minor in these positive-parity states, as they require a higher orbital angular momentum of $2$.

As mentioned in Sec.~\ref{intro}, a significant $^6\mathrm{He}+p$ resonance with angular momentum $1/2$ and positive parity in $^7$Li, only $0.23~\mathrm{MeV}$ above the threshold, has been predicted by the NCSM calculation~\cite{vorabbi2019}.
In our GCM calculation, we did not find such a $1/2^+$ state of $^7$Li near the $^6\mathrm{He}+p$ threshold.
According to our results, besides the ground and first excited states, the $^6\mathrm{He}+p$ configuration is present in the $3/2^-_3$, $3/2^-_4$, $1/2^-_3$, and $1/2^-_4$ states, along with various $^6\mathrm{Li}+n$ channels.
For $^7$Be, no physical positive-parity state is close to the $^6\mathrm{Li}+p$ threshold, which is consistent with the NCSM calculation. 
The $^6\mathrm{Li}+p$ component is significant in the $7/2^-_1$ state, which is much lower than the $^6\mathrm{Li}+p$ threshold, as well as in other states above the threshold.

\section{Summary}
\label{summary}

In the present work, we applied GCM four-body calculations to obtain the energy spectra and wave functions of the ground and excited states of $^7$Li and $^7$Be.
The results successfully reproduced the states found in experiments.
The calculated RWA show that the $\alpha+t$ configuration is dominant in the ground and first three excited states of $^7$Li, as well as the $\alpha+{^3\mathrm{He}}$ configuration in $^7$Be, indicating a pronounced preference for forming $t~(^3\mathrm{He})$ clusters in the presence of $\alpha$ in $^7\mathrm{Li}~(^7\mathrm{Be})$ within the lower energy range.
As the excitation energy increasing, especially beyond the $^6\mathrm{Li}+n$ or $^6\mathrm{Li}+p$ threshold, the $\mathrm{core}+N$ configurations become more prominent, and the $\alpha+t~(^3\mathrm{He})$ component is suppressed due to the breakup effect of $t~(^3\mathrm{He})$ cluster.
Additionally, the RWA of configurations with similar coupling channels of angular momenta exhibit analogous characteristics.
However, the method utilized in the present work can only analyze two-body cluster structures in nuclei.
The components of three-body configurations, which are closely related to the formation of deuteron, dineutron, and diproton clusters, and even four-body configurations in $^7$Li and $^7$Be will be studied in the future.

\begin{acknowledgments}
This work is supported by the National Key R$\&$D Program of China (2023YFA1606701). This work was supported in part by the National Natural Science Foundation of China under contract Nos. 12175042, 11890710, 11890714, 12047514, 12147101, and 12347106, Guangdong Major Project of Basic and Applied Basic Research No. 2020B0301030008, and China National Key R$\&$D Program No. 2022YFA1602402. This work was partially supported by the 111 Project.
\end{acknowledgments}

%
    
\end{CJK*}
\end{document}